\documentclass[1p,number,sort&compress]{elsarticle}

\journal{}

\begin{document}

%%%%%%%%%%%%%%%%%%%%%%%%%%%%%%%%%%%%%%%%%%%%%%%%%%%%%%%%%%%%%%%%%
%% FRONTMATTER
%%%%%%%%%%%%%%%%%%%%%%%%%%%%%%%%%%%%%%%%%%%%%%%%%%%%%%%%%%%%%%%%%

% title
\title{Integration of Renewable Energy Sources in Future Power
  Systems: The Role of Storage}%
% author information
\author[ne]{Stefan Weitemeyer\corref{cor1}}%
\ead{stefan.weitemeyer@next-energy.de}%
\author[ne]{David Kleinhans} \author[ne]{Thomas Vogt}%
\author[ne]{Carsten Agert}%
\cortext[cor1]{Corresponding author. NEXT ENERGY, EWE Research Centre
  for Energy Technology at the University of Oldenburg,
  Carl-von-Ossietzky-Str.~15, 26129 Oldenburg, Germany. Tel.:~+49 441
  99906 105.}%
\address[ne]{NEXT ENERGY, EWE Research Centre for Energy Technology at
  the University of Oldenburg, 26129 Oldenburg, Germany}

% abstract
\begin{abstract}
  Integrating a high share of electricity from non-dispatchable
  Renewable Energy Sources in a power supply system is a challenging
  task. One option considered in many studies dealing with prospective
  power systems is the installation of storage devices to balance the
  fluctuations in power production. However, it is not yet clear how
  soon storage devices will be needed and how the integration process
  depends on different storage parameters. Using long-term solar and
  wind energy power production data series, we present a modelling
  approach to investigate the influence of storage size and efficiency
  on the pathway towards a 100\% RES scenario. Applying our approach
  to data for Germany, we found that up to 50\% of the overall
  electricity demand can be met by an optimum combination of wind and
  solar resources without both curtailment and storage devices if the
  remaining energy is provided by sufficiently flexible power
  plants. Our findings show further that the installation of small,
  but highly efficient storage devices is already highly beneficial
  for the RES integration, while seasonal storage devices are only
  needed when more than 80\% of the electricity demand can be met by
  wind and solar energy. Our results imply that a compromise between
  the installation of additional generation capacities and storage
  capacities is required.
\end{abstract}

% keywords etc (separated by \sep)
\maketitle

%%%%%%%%%%%%%%%%%%%%%%%%%%%%%%%%%%%%%%%%%%%%%%%%%%%%%%%%%%%%%%%%%
%% BODY
%%%%%%%%%%%%%%%%%%%%%%%%%%%%%%%%%%%%%%%%%%%%%%%%%%%%%%%%%%%%%%%%%

%\linenumbers

\section{Introduction}
On the pathway towards a prospective low carbon energy system, the
share of electricity produced from Renewable Energy Sources (RES) in
the European power supply system has increased significantly over the
past years \cite{Eurostat2014}. Ongoing concerns about climate change
and the aim of many countries to become more independent from energy
imports will likely lead to a further increase in the share of RES in
the European electricity supply system \cite{Foundation2010}.

In such a system, the major share of energy would be provided by wind
and solar energy as they are considered to have the highest potential
in Europe \cite{Jacobson2011}. Due to their natural origin the
electricity produced from these sources is fluctuating strongly on
both short-term (seconds to hours) and long-term scales (months,
years) \cite{Georgilakis2008, Heide2010}. As production and
consumption in a power supply system always need to be balanced, there
is a requirement for reserve power capacities to ensure the security
of supply, in the form of either quickly adjustable back-up power
plants (operated e.g.\ on gas) or storage units \cite{Foundation2010,
  Bunger2009, Schill2013, Heide2010, Rasmussen2012}. Storages can
store surplus electricity generated when the production from RES
exceeds the demand and, hence, reduce the need for curtailment of
electricity produced from RES \cite{Hart2012}.

Already with today's European power supply system with slightly more
than 20\% of the electricity demand covered from RES
\cite{Eurostat2014}, it is debated which share of electricity produced
from fluctuating RES the current power supply system can
handle. According to a contribution by Hart et al.\ \cite{Hart2012}
the integration of RES in the power system can generally be
characterised by two phases: Up to a certain penetration of RES, all
the electricity produced from RES can be fed into the system, thus the
integration of RES scales linearly with RES capacities
\cite{Hart2012}. After a certain transition point, the electricity
production from RES occasionally exceeds the energy demand implying
the need for curtailment of RES to ensure grid stability
\cite{Hart2012}. In this second phase the integration of RES scales
less than linear with the installed capacities \cite{Hart2012}.

Another contribution investigated the effect of transmission grid
extension on this integration process \cite{Schaber2012}. The authors
showed that a powerful overlay transmission grid significantly reduced
overproduction and back-up capacity requirements
\cite{Schaber2012}. Furthermore, grid expansion was found to be also
favourable from an economic perspective over only installing more
variable renewable energy capacities \cite{Schaber2012,Fuersch2013}.

In addition to back-up power plants many studies dealing with
prospective power supply systems with a high share of RES investigate
the utilisation of storage devices to balance the fluctuations in the
electricity production from RES (see
e.g.~\cite{Bunger2009,Schill2013,Rasmussen2012,Wiese2014,Lund2009,Heide2010,Heide2011,
  Weiss2013,Andresen2014} for Europe, \cite{Elliston2012,Elliston2013}
for Australia and \cite{Budischak2012,Nelson2012} for the United
States). Some of these studies implement very detailed assumptions on
the cost for installation and operation of relevant units
\cite{Schill2013,Budischak2012, Lund2009, Elliston2012, Elliston2013,
  Nelson2012}. In order to promote a deeper understanding of the
dependencies and implications relevant for the transformation of the
power supply system, however, systematic investigations of fundamental
aspects of the integration of RES are required. This paper addresses
the impact of storages on the integration of RES in general and the
importance of their size and efficiency in particular. Both the
general approach and the results obtained for Germany are intended to
set the stage for more detailed studies on the economic aspects on
their integration and operation.

\section{Modelling storage in power systems}
A prospective power supply system based almost entirely on RES will
depend strongly on wind and solar resources and, hence, needs to deal
with their intrinsic variability. This work focusses on the
large-scale integration of RES from a meteorological perspective. For
this purpose we assume that representative data on power generation
from wind $W(t_i)$ and solar $S(t_i)$ resources and load data $L(t_i)$
is available at discrete times $t_i=i\tau$ with $1\le i\le N$, where
$\tau$ is an arbitrary but fixed time increment. Each data point here
corresponds to the accumulated energy generated or consumed in the
respective time lag $\tau$.\footnote{In simulations in this work
  typically $\tau=1\ \mathrm{h}$ is used. For reasonable conclusions
  regarding the required storage size, the time increment $\tau$ needs
  to be sufficiently small, since relevant effects might disappear at
  larger time scales.} It is assumed that the data is corrected for
systematic changes during the period of analysis.

The resource data can either stem from measurements on existing
systems or, as in the case investigated in more detail in section
\ref{sec:application-germany}, from meteorological simulations.  In
order to ease a scaling of the generation data for the investigation
of different installed capacities the generation data is normalised
and expressed in units of the average electricity demand in the
respective observation intervals. With $\langle
X(t)\rangle_t:=N^{-1}\sum_{i=1}^NX(t_i)$ we define normalised data
sets $w$ and $s$ as
\begin{equation}
  w(t):=\frac{W(t)}{\langle W(t)\rangle_t} \cdot \langle L(t)\rangle_t,\quad  
  s(t):=\frac{S(t)}{\langle S(t)\rangle_t} \cdot \langle L(t)\rangle_t.% \nonumber
\end{equation}

The production potential is then put into relation to the
corresponding load data. A general form of the mismatch in generation
from RES and energy demand at time $t$ can be defined as
\begin{equation}
  \Delta_{\alpha,\gamma}(t):=\gamma \left(\alpha w(t)+(1-\alpha)s(t)\right)-L(t)\quad.
\end{equation}
Here, $\gamma\alpha$ and $\gamma(1-\alpha)$ render the respective
shares of wind and solar power generation of the gross electricity
demand. $\gamma$ determines the total electricity produced from RES
and is termed the average renewable energy power generation factor
(cf. \cite{Heide2011}).

In order to study the role of storage devices for the integration of
RES, we choose the following procedure: first, we investigate which
share of electricity demand can be met by RES if no storage devices
are present. Second, we add an infinitely large storage device with
round-trip efficiency $\eta$ to the system, and third, we alter the
storage device to one with limited size $H^{\max}$.

In the first case without any storages, the energy production from RES
needs to be curtailed in periods of overproduction
($\Delta_{\alpha,\gamma}(t)>0$), whereas negative mismatches
($\Delta_{\alpha,\gamma}(t)<0$) need to be balanced by back-up power
plants. The total amount of curtailed energy in multiples of the total
demand is in this case determined by the overproduction function
$O_0(\alpha,\gamma)$,
\begin{equation}
  O_0(\alpha,\gamma)=\frac{\left \langle
      \max[0,\Delta_{\alpha,\gamma}(t)]\right\rangle_t}{\left\langle L(t)\right\rangle_t}\quad.
  \label{eq:O_0}
\end{equation}

The share of energy demand met by wind or solar energy after
curtailment for a certain configuration $\alpha$ and $\gamma$, which
we will call renewable integration function $RE_0(\alpha, \gamma)$, is
then calculated as
\begin{equation}
  RE_0(\alpha,\gamma)=\frac{\left\langle \gamma \left(\alpha w(t)+(1-\alpha)s(t)\right)\right\rangle_t  - O_0(\alpha,\gamma)\left\langle L(t)\right\rangle_t}{\left\langle L(t)\right\rangle_t}=\gamma - O_0(\alpha, \gamma)\quad.
  \label{eq:RE_0}
\end{equation}
A scenario without any contribution from RES (0\% RES scenario)
consequently results in $RE_0(\alpha,\gamma)=0$. By means of
eq.~(\ref{eq:RE_0}) scenarios can then be categorized with respect to
their contribution from RES. Since by construction
$O_0(\alpha,\gamma)\ge \gamma-1$ the renewable integration function
$RE_0(\alpha,\gamma)$ has a maximum of $1$, which is realised when all
demand is provided by RES (100\% RES scenario).

Taking, secondly, also into account storages, this approach can be
generalised. For storage of sufficient size boundary effects can be
neglected. Then it is sufficient to take into account that parts of
the overproduction can be fed into the system again. If we assume
fully flexible and infinitely large storages with no self-discharging
and with a round-trip efficiency $\eta$, the share of energy demand
met by wind and solar energy is defined as the renewable integration
function $RE_\infty^\eta(\alpha,\gamma)$,
\begin{equation}
  RE_\infty^\eta(\alpha,\gamma)=\gamma - \max[\gamma-1,(1-\eta)O_0(\alpha, \gamma)]=\gamma - \max[\gamma-1,O_{\infty}(\alpha, \gamma)]\quad.
  \label{eq:RE_infty}
\end{equation}
In this definition the $\max$ function guarantees that the electricity
directly produced from RES plus the electricity re-injected from the
storages does not exceed the total demand. This becomes relevant in
particular at large $\gamma$, where one would obtain
$RE_\infty^\eta(\alpha,\gamma)>1$ otherwise.  Note that with $\eta=0$
this equation also includes the case without any storages
(i.e. $RE_\infty^{\eta=0}(\alpha,\gamma)=RE_0(\alpha,\gamma)$).

Thirdly, we address the most general case, the integration of RES with
limited storage capacities of size $H^{\max}$ (in units of $\langle
L(t)\rangle_t$). For a given wind share $\alpha$ and given average
renewable energy power generation factor $\gamma$, the storage time
series $H_{\alpha,\gamma}^{\eta}(t)$ describing the energy available
to the grid is defined iteratively as

\begin{equation}
  H_{\alpha,\gamma}^{\eta}(t)=
  \left\{\begin{array}{ll}
      \mbox{if }\Delta_{\alpha,\gamma}(t) \ge 0:\\
      \mbox{  }\min[H^{\max},H_{\alpha,\gamma}^{\eta}(t-\tau)+\eta \Delta_{\alpha,\gamma}(t)]\\
      \mbox{if }\Delta_{\alpha,\gamma}(t) < 0:\\
      \mbox{  }\max[0, H_{\alpha,\gamma}^{\eta}(t-\tau)+\Delta_{\alpha,\gamma}(t)]
    \end{array}\right.
\end{equation}
with $\eta$ being the round-trip efficiency of the fully flexible
storage. This expression is evaluated at integer multiples of $\tau$,
with $\tau$ being the fixed time increment of the time series as
defined earlier in this section. The initial charging level of the
storage $H_{\alpha,\gamma}^{\eta}(t=0)$ has to be specified when the
approach is applied to actual data (cf.\ sec.~\ref{ssec:DataGermany}).

In this case, the total amount of unusable energy (due to curtailment
and efficiency losses) in multiples of the total demand is determined
by the overproduction function
\begin{equation}
  O_H^{\eta}(\alpha,\gamma)=\frac{\left\langle\max[0,\Delta_{\alpha,\gamma}(t)-\left(H_{\alpha,\gamma}^{\eta}(t)-H_{\alpha,\gamma}^{\eta}(t-\tau)\right)]\right\rangle_t}{\langle L(t)\rangle_t}\quad.
\end{equation}
This expression merges into the respective expressions $O_0$ and
$O_{\infty}$ as defined in equations (\ref{eq:O_0}) and
(\ref{eq:RE_infty}) for the respective assumptions $\eta=0$ and
$H^{max}\gg H_{\alpha,\gamma}^{\eta}(t=0)\gg 0$.

The share of energy demand met by RES with a storage of size
$H^{\max}$ available in the system is then defined as
\begin{equation}
  RE_H^{\eta}(\alpha,\gamma)=\gamma-\max[\gamma-1,O_H^{\eta}(\alpha,\gamma)]\quad.
  \label{eq:RE_H}
\end{equation}

\section{\label{sec:application-germany}Application to Germany}
The methods developed in the previous section are now applied to
specific data in order to study the role of energy storage devices for
the integration of RES in future power systems. Due to the
availability of resource and demand data as well as a RES penetration
of over 20\% in its electricity system \cite{Eurostat2014}, Germany is
chosen for this purpose.

\subsection{Data used for calculations}
\label{ssec:DataGermany}
The production from RES in Germany is estimated using long-term solar
($S(t)$) and wind energy ($W(t)$) power output data series with hourly
resolution ($\tau=1\ \mathrm{h}$) spanning eight years from 2000 to
2007 and based on reanalysis data (for details we refer to
\cite{Heide2010}). We only handle aggregated time series and do not
consider limitations and effects of the national grid. Exports and
imports of electricity are not considered in this work (see also
section \ref{sec:impl-size-simul} for the potential implications). The
production from the fluctuating RES is put into relation to the demand
load time series $L(t)$, which is available from ENTSO-E.\footnote{The
  respective data series for both production and load have also been
  used in previous publications by other authors,
  \cite{Heide2010,Heide2011,Rasmussen2012,Rodriguez2013,
    Andresen2014}.} All time series are normalised to the average
load, hence a power generation factor $\gamma=1.0$ corresponds to
scenarios where the total electricity producible by wind and solar
resources in the eight-year period is identical to the overall load
during this time. This way, our results do not depend explicitly on
the absolute power generation capacities.

With the data for Germany we now evaluate the renewable integration
functions (eq.~(\ref{eq:RE_0}), (\ref{eq:RE_infty}) and
(\ref{eq:RE_H})) and investigate their dependence on the wind share
$\alpha$ and the generation factor $\gamma$. The storage size
$H^{\max}$ is chosen to be equivalent to 0, 2, 4, 6, 8, 10, 12, 24,
36, 48, 72, 168, 360, 720, 1440 average load hours (av.l.h.); for the
case presented in this work (Germany) holds
1~av.l.h.$\hat{=}$54.2~GWh. We checked if the initial charging level
of the storage $H_{\alpha,\gamma}^{\eta}(t=0)$ has an influence on the
results and we found that the results only change by a maximum of
0.1\% in all cases considered in this work. Hence, the initial
charging level of the storage was set to
$H_{\alpha,\gamma}^{\eta}(t=0)=0.5H^{\max}$ without further
discussion.

\subsection{Results}
\label{ssec:Results}
\begin{figure}
  \centering
  \includegraphics[width=0.75\columnwidth]{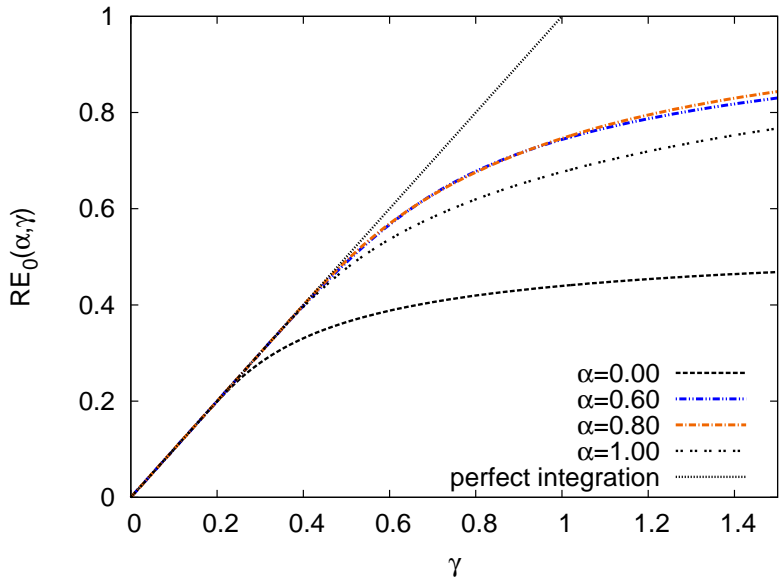}
  \caption{Renewable integration function $RE_0(\alpha,\gamma)$ for
    the case without any storage and selected wind shares $\alpha$}
  \label{fig:sampleRE0}
\end{figure}

The renewable integration function $RE_0(\alpha,\gamma)$ for the case
of no storage and selected wind shares $\alpha$ is shown in figure
\ref{fig:sampleRE0}. For $\gamma < 0.2$, there is no overproduction,
hence all curves rise linearly in this section. With increasing power
generation factor $\gamma$ overproduction occurs more and more
frequently and the curves bend down due to curtailment. The transition
point between the linear regime and the curtailment regime depends on
the wind share $\alpha$. In a solar-only scenario ($\alpha=0.00$) the
effects of curtailment become relevant already at $\gamma= 0.2$.
Previous investigations have shown that the overproduction is mainly
due to the solar production peak around noon. The curves with $0.60
\le \alpha \le 0.80$ are the topmost, in accordance with
\cite{Heide2010} the overproduction is least with a mix of solar and
wind energy in this regime. Unless otherwise noted, we will use a wind
share of $\alpha=0.60$ for the following results.  For this wind
share, the transition occurs at about $\gamma=0.5$. However, sooner or
later the renewable integration functions flatten out significantly
independently of the wind share $\alpha$, implying that a massive
installation of additional production capacities would be needed to
increase the integration of RES up to 100\%. Thus, after a certain
penetration of RES, the installation of storage capacities is likely
to be worthwhile.

If storages are available, parts of the otherwise curtailed energy can
be fed into the system again, whereas the amount of reusable energy
depends on storage size and round-trip efficiency (cf.\
eq.~(\ref{eq:RE_infty})). Hence, the renewable integration function
$RE_\infty^\eta(\alpha,\gamma)$ after the transition point increases
compared to the case without any storage.

For an unlimited storage without any losses ($\eta=1$), all
electricity produced during periods of overproduction can be fed into
the system again in times of underproduction. Hence, the transition
point between the linear regime and the curtailment regime moves to
$\gamma=1$.  We estimated the storage size required to have the same
properties as an infinitely large storage device by calculating the
spread of the cumulative sum of the mismatch function
$\Delta_{\alpha,\gamma}(t)$.  In a scenario for Germany with a wind
share of $\alpha=0.60$ and an average renewable energy power
generation factor of $\gamma=1.00$, a loss-free storage device
($\eta=1$) would need to have a size in the order of 80 TWh. This is
orders of magnitudes higher than today's storage capacities in Germany
(39~GWh for pumped-hydro storage according to \cite{Eurelectric2011})
and even higher than Europe's total hydrogen storage potential in salt
caverns (32~TWh according to \cite{Perez2013}).  As can be seen from
equation (\ref{eq:RE_infty}) with unlimited storage capacities, a
100\% RES scenario would always be possible. In order to balance the
losses of the storage, however, additional generation capacities would
need to be installed. The extent of these overcapacities depends on
the round-trip and increases e.g.\ to $\gamma\approx 2.0$ for
$\eta=0.1$.

\begin{figure}
  \centering \includegraphics[width=0.75\columnwidth]{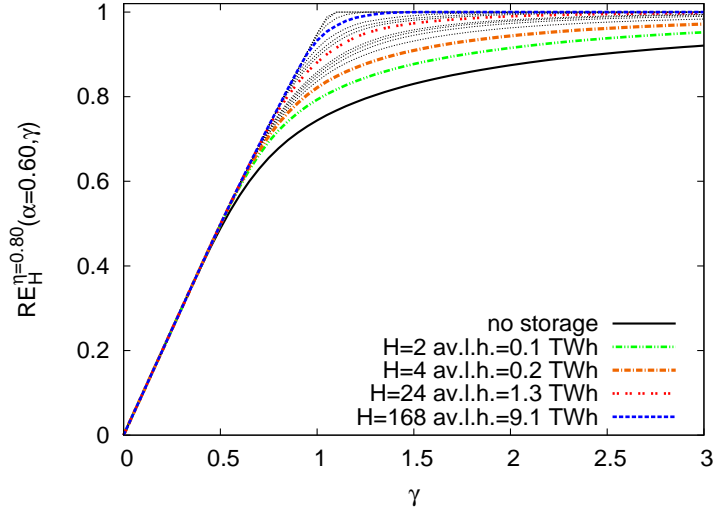}
  \caption{Renewable integration function
    $RE_H^{\eta=0.8}(\alpha=0.60,\gamma)$ for different storage sizes
    $H^{\max}$ and storage round-trip efficiency $\eta=0.8$. Dashed
    lines represent additional intermediate storage sizes (cf.\
    sec.~\ref{ssec:DataGermany})}
  \label{fig:REofHmax}
\end{figure}

We will now proceed to storages of limited
sizes. Figure~\ref{fig:REofHmax} shows the renewable integration
function $RE_H^{\eta=0.8}(\alpha=0.60,\gamma)$ for different storage
sizes $H^{\max}$ and storage round-trip efficiency $\eta=0.8$ (which
is a typical value for pumped-hydro storage facilities). The wind
share $\alpha$ has been fixed to $\alpha=0.60$, which previously was
found to be close to the mix at which overproduction occurs last
(cf. fig.\ \ref{fig:sampleRE0}). In figure \ref{fig:REofHmax} the
transition between the previously discussed cases of no storage and
unlimited storage capacities can be observed. Furthermore, one can see
that already a storage with a capacity of only
$H^{\max}=2~\mbox{av.l.h.}\hat{=}0.1~\mbox{TWh}$ significantly
increases the integration of RES. This can be understood from the fact
that already the availability of small storage devices increases the
integration of solar energy significantly and extents its usability to
the evenings. The effect of storages decreases with their size.

\begin{figure}
  \centering
  \includegraphics[width=0.75\columnwidth]{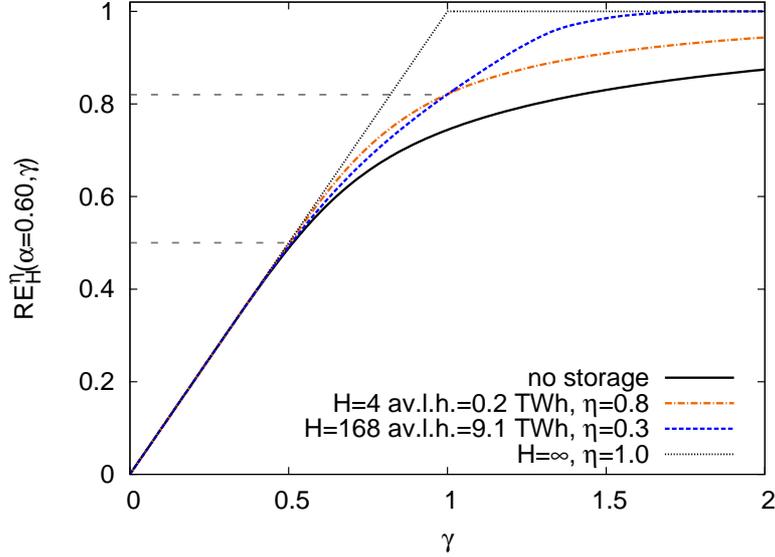}
  \caption{Renewable integration function
    $RE_H^{\eta}(\alpha=0.60,\gamma)$ for two different storage
    classes: Comparison between small, highly efficient storages and
    large, less efficient ones. Black lines represent the case of no
    storage as well as infinitely large loss-free storages.}
  \label{fig:DE_comp_twostorages}
\end{figure}

Finally, let us come back to the question raised in the introduction,
how soon will we need storage devices when increasing the share of
RES. Figure \ref{fig:DE_comp_twostorages} shows the evolution of the
renewable integration function $RE_H^{\eta}(\alpha=0.60,\gamma)$ for
two different storage classes, first, small and highly efficient
storages ($H^{\max}=4~\mbox{av.l.h.}\hat{=}0.2~\mbox{TWh}$,
$\eta$=0.8) such as e.g.\ pumped-hydro storage facilities and second,
large storages with reduced efficiency
($H^{\max}=168~\mbox{av.l.h.}\hat{=}9.1~\mbox{TWh}$, $\eta$=0.3), as
it e.g.\ can be assumed for seasonal storages based on synthetic
hydrogen. For comparison, the evolution of the renewable integration
function without storage $RE_0(\alpha=0.60,\gamma)$ and unlimited
storage without any losses $RE_{\infty}^{\eta=1}(\alpha=0.60,\gamma)$
are also plotted. One can see that up to a share of about 50\% of the
energy demand met by RES the curves do not exhibit significant
differences. That is, no storage would be needed until this point,
provided that the remaining load can be met by fully flexible power
plants as assumed in this approach. For a share of about 50-80\% of
the energy demand met by wind and solar energy, a small but efficient
storage can achieve a better integration of RES than a large but less
efficient storage device. Only for higher shares of RES their
integration is higher with large seasonal storage. With this storage a
100\% RES scenario would be possible with a power generation factor
$\gamma$ in the order of about $\gamma$=1.5. For a better discussion
of the results and a comparison with more recent data, corresponding
calculations were done using a different data source spanning the
years from 2006 to 2012. Apart from minor quantitative differences,
both data sets lead to the comparable results that up to 50\% of the
electricity demand could be met without storage; and small but highly
efficient storage devices should be favoured over large but less
efficient storage devices to reach a share of about 80\% of the
electricity demand being met by RES. Systematic differences between
the observation periods 2000-2008 and 2006-2012 were not found. For
details we refer to \ref{sec:appendix-secdata}.

These results mean that at the beginning of the RES integration
process rather small and highly efficient storage devices are
sufficient if combined with flexible fossil power plants. Only when it
comes to integrating very high shares of RES, hence having a system of
almost 100\% RES, seasonal storage devices are needed. Alternatively,
overcapacities could be installed to reach a 100\% RES system while
requiring less storage capacity. Hence, a compromise between the
installation of overcapacities and the installation of storage
capacities has to be found.

\section{Further aspects of storage integration}
In the previous section, we focussed on the conceptual question on how
to include different storage parameters when studying the integration
of RES in prospective power supply systems. We now take a look at
three further aspects closely related to the previous results, namely
the dependence of the storage requirements on the mix between wind and
solar resources, the economic impact of our previous findings and the
implications of the size of the investigated region.

\subsection{Storage and the optimal mix}
\begin{figure}
  \centering \includegraphics[width=0.75\columnwidth]{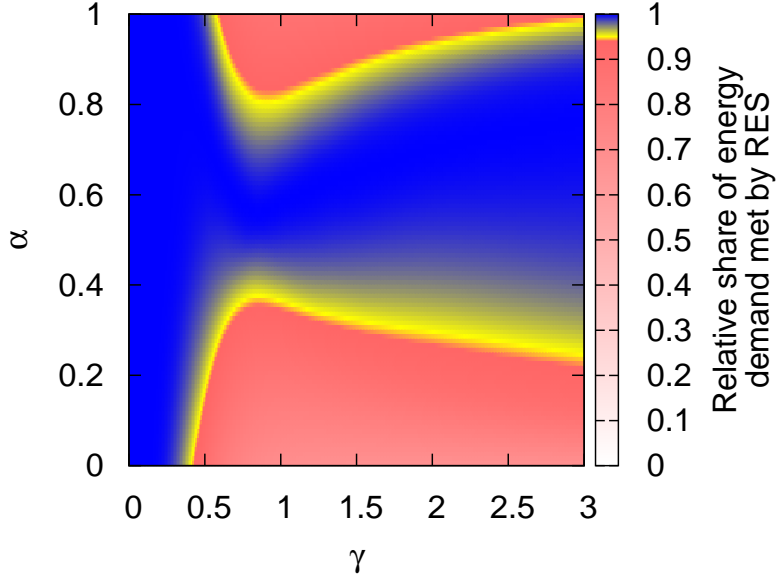}
  \caption{Dependence of the wind share $\alpha$ on maximum curve for
    storage with $H^{\max}=4~\mbox{av.l.h.}\hat{=}0.2~\mbox{TWh}$ and
    round-trip efficiency $\eta=0.8$.  For each power generation
    factor $\gamma$, the figure shows how close the value
    $RE_H^{\eta=0.8}(\alpha,\gamma)$ for each wind share $\alpha$ is
    compared to the maximum value for this power generation factor
    $\gamma$. The blue area (inner part of the figure) represents
    parameters where the integration is at least 95\% of the maximum
    value.}
  \label{fig:distr_H4}
\end{figure}

We have shown above that the penetration of RES at which
overproduction and hence possible curtailment starts to take place
depends on the mix between solar and wind energy (cf.\
fig.~\ref{fig:sampleRE0}). In the subsequent results, the wind share
was fixed to $\alpha = 0.6$, which was found to be in the range of the
optimal mix for minimum storage capacities. To investigate the
influence of the mix in more detail, figure \ref{fig:distr_H4} shows
for each generation factor $\gamma$ the ratio of the integration for
each wind share $\alpha$ to the integration of the wind share $\alpha$
with the best integration for this generation factor $\gamma$. Here, a
particular storage size of
$H^{\max}=4~\mbox{av.l.h.}\hat{=}0.2~\mbox{TWh}$ with a round-trip
efficiency $\eta=0.8$ is considered. The blue area represents
parameters for which the integration is at least 95$\%$ of the maximum
value. In accordance with figure \ref{fig:sampleRE0} for $\gamma\le
0.3$ the integration does not depend on the wind share $\alpha$. For
higher values of $\gamma$ (corresponding to higher RES capacities),
the optimal wind share $\alpha$ lies in the range of
$0.5<\alpha<0.65$. Any wind share $\alpha$ is this range leads to a
curve which differs less than 5\% from the integration at the best
wind share $\alpha$. With increasing generation factor $\gamma$, the
blue area widens towards higher and lower values of $\alpha$, hence
for larger installed capacities the integration becomes less sensitive
to the wind share $\alpha$. Overall, these results imply that the mix
between wind and solar capacities is especially important during the
transition process ($0.3\leq \gamma \leq 1.5$) to best integrate
RES. For lower and higher penetrations of RES, however, a broad range
of wind shares $\alpha$ leads to an optimal integration of RES.

\subsection{Economic impact of storage integration}
The results shown in this work are based mainly on meteorological
aspects and the resulting fluctuations in the power production from
solar and wind resources as well as the load patterns in the current
power supply system. We eventually take a look at the economic impact
of our previous findings. As shown in figures
\ref{fig:sampleRE0}-\ref{fig:DE_comp_twostorages}, there is a linear
regime at the beginning of the integration of renewable energy
sources, implying all electricity produced by RES can be integrated
completely into the electricity grid. Provided that the remaining
back-up power plants are fully flexible, the installation of storage
devices is economically not directly profitable in this regime. Once
curtailment sets in implying that the curves for the respective
renewable integration functions bend downwards, there are different
pathways to increase the share of energy demand met by RES: One option
is to install storage capacities (cf.\
fig.~\ref{fig:REofHmax}). Alternatively, the same share can be
achieved by constructing additional power generation capacities, hence
increase the generation factor $\gamma$. If the latter option is
chosen, however, the system operator (referring to e.g.\ the grid
operator, the government, the overall society or anyone responsible
for the stable supply of electricity) has to find a way to make the
new investments profitable for an investor.

\begin{figure}
  \centering
  \includegraphics[width=0.75\columnwidth]{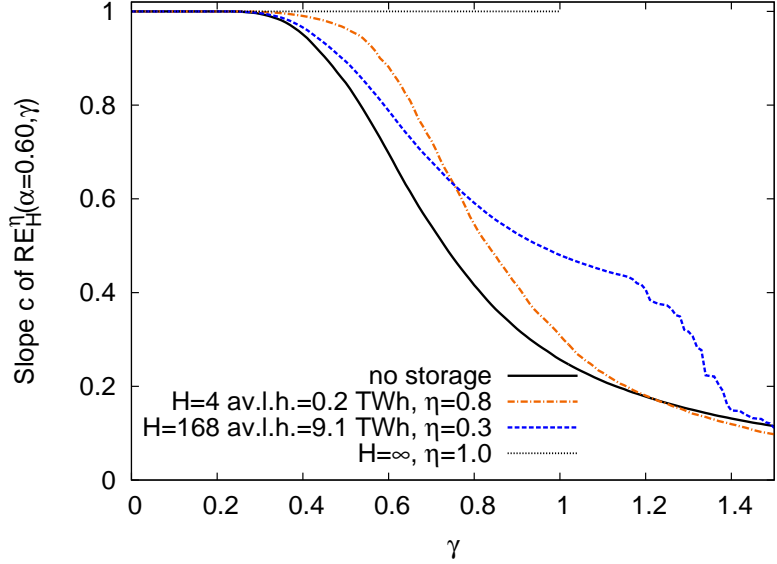}
  \caption{Slope $c$ of the renewable integration function
    $RE_H^{\eta}(\alpha,\gamma)$ as a function of the renewable energy
    power generation factor $\gamma$ for two different storage classes
    as well as the case of no storage and infinitely large storages.}
  \label{fig:compareslope}
\end{figure}

Let us look again at the case of Germany shown before in figure
\ref{fig:DE_comp_twostorages}. Figure~\ref{fig:compareslope} shows the
slope, denoted $c$, of the renewable integration function
$RE_H^{\eta}(\alpha,\gamma)$ as a function of the generation factor
$\gamma$ for the different cases discussed previously. For scenarios
with a share of energy demand being met by RES above 80\%, hence
$\gamma>1.00$ (cf.\ fig.~\ref{fig:DE_comp_twostorages}), the slope $c$
is at a value of $c\le0.5$. This would mean that in this simplified
scenario less than half of the additionally producible electricity
could be fed into the system.\footnote{The surplus energy could be
  transferred to the heat or transport sector using additional
  conversion capacities; investigating this option in detail is beyond
  the scope of this work though.}  Either this would seriously affect
the return on investment of additional conversion facilities or
alternatively all previously installed RES capacities would have to
slightly curtail their electricity production. Both effects would have
a high impact on decisions regarding the investment into additional
generation capacities.

In addition to the capacity of the storages, from an economic point of
view also the required converter power for charging and discharging
are relevant. The model presented in this work for reasons of
simplicity does not incorporate any restrictions in the converter
power and, hence, does not enforce an economic utilisation of the
converters. In the scenario with a wind share $\alpha=0.60$ and a
power generation factor $\gamma=1.0$ the realised power of a storage
device with $H^{\max}=4~\mbox{av.l.h.}$ and round-trip efficiency
$\eta=0.8$ e.g.\ was about $1.23\langle L(t)\rangle_t\hat{=}67\,
\mathrm{GW}$ (95\% quantile: $0.81\langle L(t)\rangle_t\hat{=}44 \,
\mathrm{GW}$) for discharging and $1.69\langle
L(t)\rangle_t\hat{=}92\, \mathrm{GW}$ (95\% quantile: $1.05\langle
L(t)\rangle_t\hat{=}57\, \mathrm{GW}$) for charging,
respectively. These values are much higher than the converter power
requirements found in a similar studies (see e.g.\ \cite{Weiss2013a}),
which do take into account economic aspects in the operation of
storages and backup power. Since the utilization of converters in the
present contribution is not optimized, derived parameters such as
capacity factors for the storage devices are not meaningful and cannot
be included in an evaluation of economic aspects as e.g. done by
\cite{Anagnostopoulos2012}.

Limiting the admissible converter power of the storage devices in the
simulations presented here in particular for smaller storage sizes
does not necessarily imply significant changes in the results for the
renewable integration function, since the limits initially only change
the interaction between storage devices and back-up power plants
(restricting e.g.\ the power of the storage with size
$H^{\max}=4~\mbox{av.l.h.}$ to their 95\% quantile did not have any
significant effect). A detailed and systematic discussion of the
impact of converter power, which would be required for full evaluation
of economical aspects of the integration of storages, however, is
beyond the scope of this manuscript and will instead be subject of
future work.

\subsection{\label{sec:impl-size-simul}Implications of size of the
  simulation domain}
The model developed in this work focusses on the role of storage
devices for the integration of RES while treating the simulated domain
as a copper plate.  Grid limitations are not considered. Instead we
assume that the production from RES and the demand can be balanced
without constraints throughout the entire simulation
domain. Consequently the integration of renewables is generally
promoted with increasing size of the simulation domain
\cite{Steinke2013}, since fluctuations in production and consumption
decorrelate with increasing scales. By expanding our model to a
European scale\footnote{The analysis includes the former UCTE member
  countries as well as Ireland, United Kingdom, Norway, Sweden,
  Finland and Denmark. For details we refer to \cite{Heide2010}.} the
renewable integration function for a storage device with size
$H^{\max}=2~\mbox{av.l.h.}$ for instance increases from
$RE_H^{\eta=0.8}(\alpha=0.60,\gamma=1.50)\approx0.88$ for Germany to
$RE_H^{\eta=0.8}(\alpha=0.60,\gamma=1.50)\approx0.94$. In this
respect, the effect of imports and exports to foreign grids could be
regarded as a virtual storage. However, the positive effects of
enlarging the investigated region are restricted by the capacity of
the grid (in particular but not limited to interconnectors)
\cite{Rodriguez2013}. In practise these restrictions would need to be
solved either by expansion and enforcement of distribution and
transmission grids or by the installation of additional storages to
overcome local bottlenecks. For this reason the results derived from
our approach for large-scale systems with rather mature transmission
grids (such as e.g.\ Germany and Europe) exhibit lower bounds for the
actual storage demand.

\section{Conclusions}
The transformation of the European power supply system to one based on
Renewable Energy Sources (RES) is a challenging task -- yet it is
achievable. In order to balance the fluctuations in the power
production from wind and solar energy, the installation of storage
capacities will likely be required. The storage modelling approach
developed in this work allows to systematically study the integration
process of RES in a power system depending on the round-trip
efficiency and the size of the storage.

Applying our approach to data for Germany, we found that up to 50\% of
the electricity demand could be met by RES without storage -- provided
that an optimal mix between wind and solar power generation is chosen
and the remaining power plants are fully flexible. This result is in
line with recently published case studies for Germany \cite{Weiss2013}
and Denmark \cite{Andresen2014}. The required flexibility is, however,
currently only the case for parts of the German generation
portfolio. In a scenario with these flexible back-up power plants,
though, small but highly efficient storage devices should be favoured
over large but less efficient (seasonal) storage devices to reach a
share of about 80\% of the electricity demand being met by
RES. Eventually a balance between the installation of additional
generation capacities and of storage capacities has to be found. In
this context also the required power of the storage converters needs
to be taken into account, which is economically relevant but beyond
the scope of the current modelling approach.

The overall transformation process involves many different
aspects. Even if only technological parameters are considered, current
energy system models tend to get very complex. By focussing our work
on two key parameters of modelling energy storage devices, we are able
to systematically study the role of storage devices for the
integration process of RES. Our approach and its findings can now be
used in upcoming modelling approaches of future power systems. By
considering more and more elements that very likely have an impact on
a future power supply system, detailed scenarios can be investigated
to achieve the goal of a cost-effective and stable supply of
electricity based on RES.

\section*{Acknowledgements}
The authors kindly thank Martin Greiner (Aarhus University, Denmark)
for providing the data used for the analysis in the main part of the
manuscript and L\"uder von Bremen (ForWind Centre for Wind Energy
Research, University of Oldenburg, Germany) for an additional data set
analyzed and discussed in the appendix. The latter data set originates
from the joint project RESTORE 2050 (funding code 03SF0439A) funded by
the German Federal Ministry of Education and Research through the
funding initiative Energy Storage. We kindly acknowledge two anonymous
reviewers and Gordon Taylor for helpful comments on an earlier version
of this manuscript. Financial support of the first author is provided
by the Lower Saxony Ministry for Science and Culture (MWK) through a
PhD scholarship.

\appendix

\section{\label{sec:appendix-secdata}Comparison of results using
  different data sets}

For a better discussion of the results presented in section
\ref{sec:application-germany} (the corresponding data set henceforth
is referred to as data set A) we applied our model on a second data
set for Germany spanning the years 2006-2012 also with $\tau =
1\mathrm{h}$ (data set B). As in data set A, the wind and solar power
generation time series of data set B are based on weather data with a
high spatial and temporal resolution. The time-dependent availability
of resources were estimated from Meteosat satellite measurements and
from a numerical weather prediction model. Within each grid cell,
sub-models for PV and wind power generation capacities transfer the
weather data to power generation, whereupon the exact spatial
distribution of the generation capacities is based on a combination of
statistical and empirical methods. Eventually, the power generation
time series are acquired by a summation over all grid cells. The
corresponding load data were obtained from ENTSO-E.

The two data sets A and B differ not only in the period of observation
but also in the underlying data sources: The solar power output data
is now based on satellite measurements instead of simulations with a
mesoscale model as in data set A. PV generation simulated from
satellite measurements generally reproduce the local weather
conditions more accurately and therefore show a higher variability in
time, even if accumulated to the country scale.  Furthermore,
parameters like the spatial distribution of the generation capacities
within the country are slightly different.

\begin{figure}
  \centering
  \includegraphics[width=0.75\columnwidth]{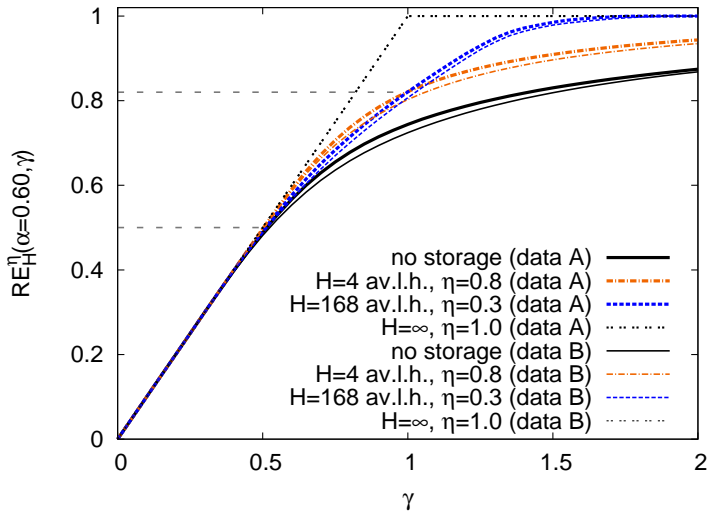}
  \caption{Comparison between two different data sets: Renewable
    integration function $RE_H^{\eta}(\alpha=0.60,\gamma)$ for
    different storage classes (cf.\
    fig.~\ref{fig:DE_comp_twostorages}) using two different data sets:
    Thick lines represent the original data set (A), results derived
    from the new, second data set (B) are presented in thick lines.}
  \label{fig:DE_comp_twodatasets}
\end{figure}

The analysis from the core part of the manuscript was now repeated on
data set B. Figure \ref{fig:DE_comp_twodatasets} exhibits a comparison
of the results with the initial results for data set A as shown in
fig.\ \ref{fig:DE_comp_twostorages}. While the integration function
for all relevant cases (limited storage or no storage at all) for data
set B is slightly lower than for data set A the gross effect of the
installation of storages and the interplay between storages of
different size is similar: Also for data set B the results suggest
that initially (and until $\gamma\approx 1$) the effect of the
installation of small and efficient storage devices is beneficial to
the installation of large and less-efficient storage capacities.

Both data sets cover a comparable period of time, which is important
to obtain a reasonable representation of regularly and extreme weather
phenomena. The smaller values for the integration function for data
set B can be attributed to the higher variability of the solar
production data, which are not correlated with the demand and
therefore have a negative impact. From this analysis and further
investigations performed we cannot attribute the differences to any
systematic differences between the years covered by the respective
data sets. This is not astonishing since the underlying data (weather
data and load data) except for the PV data do not differ significantly
between data sets A and B.

In summary, the general results are the same for both data sets A and
B: Up to 50\% of the electricity demand could be met without storage;
and small but highly efficient storage devices should be favoured over
large but less efficient (seasonal) storage devices to reach a share
of about 80\% of the electricity demand being met by RES.

%%%%%%%%%%%%%%%%%%%%%%%%%%%%%%%%%%%%%%%%%%%%%%%%%%%%%%%%%%%%%%%%%
%% REFERENCES, ETC.
%%%%%%%%%%%%%%%%%%%%%%%%%%%%%%%%%%%%%%%%%%%%%%%%%%%%%%%%%%%%%%%%%

\end{document}